\documentclass[twocolumn,twoside,hypertext]{revtex4}
\usepackage{graphicx}
\usepackage{fancyhdr}
\usepackage{amsmath} 
\usepackage{amssymb}
\usepackage{amsfonts}
\usepackage{upgreek} 
\usepackage{bm}
\usepackage{ifthen} 
\newboolean{uprightparticles}
\setboolean{uprightparticles}{false} 
\usepackage{xspace} 
\usepackage{upgreek}

\def\lhcb   {\mbox{LHCb}\xspace}

\def\babar  {\mbox{BaBar}\xspace}
\def\belle  {\mbox{Belle}\xspace}
\def\belletwo {\mbox{Belle~II}\xspace}
\def\besiii {\mbox{BESIII}\xspace}
\def\cleo   {\mbox{CLEO}\xspace}

\def\lhc    {\mbox{LHC}\xspace}

\def\MagUp {\mbox{\em Mag\kern -0.05em Up}\xspace}

\ifthenelse{\boolean{uprightparticles}}%
{

 \def\Ppi         {\ensuremath{\uppi}\xspace}                 
                  
 \def\Prho        {\ensuremath{\uprho}\xspace}

 \def\Ppsi        {\ensuremath{\uppsi}\xspace}

 \def\PDelta      {\ensuremath{\Delta}\xspace}                 
 \def\PXi         {\ensuremath{\Xi}\xspace}                 
 \def\PLambda     {\ensuremath{\Lambda}\xspace}                 
 \def\PSigma      {\ensuremath{\Sigma}\xspace}                 
 \def\POmega      {\ensuremath{\Omega}\xspace}                 
 \def\PUpsilon    {\ensuremath{\Upsilon}\xspace}

 \def\PB      {\ensuremath{\mathrm{B}}\xspace}                 
                  
 \def\PD      {\ensuremath{\mathrm{D}}\xspace}

 \def\PJ      {\ensuremath{\mathrm{J}}\xspace}                 
 \def\PK      {\ensuremath{\mathrm{K}}\xspace}

 \def\Pb      {\ensuremath{\mathrm{b}}\xspace}                 
 \def\Pc      {\ensuremath{\mathrm{c}}\xspace}                 
 \def\Pd      {\ensuremath{\mathrm{d}}\xspace}                 
 \def\Pe      {\ensuremath{\mathrm{e}}\xspace}

 \def\Pi      {\ensuremath{\mathrm{i}}\xspace}

 \def\Ps      {\ensuremath{\mathrm{s}}\xspace}                 
 \def\Pt      {\ensuremath{\mathrm{t}}\xspace}                 
 \def\Pu      {\ensuremath{\mathrm{u}}\xspace}

 \def\thebaroffset{0.0em}
}
{

 \def\Ppi         {\ensuremath{\pi}\xspace}                 
                  
 \def\Prho        {\ensuremath{\rho}\xspace}

 \def\Ppsi        {\ensuremath{\psi}\xspace}                 
                  
 \mathchardef\PDelta="7101
 \mathchardef\PXi="7104
 \mathchardef\PLambda="7103
 \mathchardef\PSigma="7106
 \mathchardef\POmega="710A
 \mathchardef\PUpsilon="7107
                  
 \def\PB      {\ensuremath{B}\xspace}                 
                  
 \def\PD      {\ensuremath{D}\xspace}

 \def\PJ      {\ensuremath{J}\xspace}                 
 \def\PK      {\ensuremath{K}\xspace}

 \def\Pb      {\ensuremath{b}\xspace}                 
 \def\Pc      {\ensuremath{c}\xspace}                 
 \def\Pd      {\ensuremath{d}\xspace}                 
 \def\Pe      {\ensuremath{e}\xspace}

 \def\Pi      {\ensuremath{i}\xspace}

 \def\Ps      {\ensuremath{s}\xspace}                 
 \def\Pt      {\ensuremath{t}\xspace}                 
 \def\Pu      {\ensuremath{u}\xspace}

 \def\thebaroffset{0.18em}
}
\newcommand{\offsetoverline}[2][\thebaroffset]{\kern #1\overline{\kern -#1 #2}}%

\makeatletter
 \newcommand{\miniscule}{\@setfontsize\miniscule{4}{5}}
\makeatother

\DeclareRobustCommand{\optbar}[1]{\shortstack{{\miniscule (\rule[.5ex]{1.25em}{.18mm})}
  \\ [-.7ex] $#1$}}

\def\en         {{\ensuremath{\Pe^-}}\xspace}   
\def\ep         {{\ensuremath{\Pe^+}}\xspace}

\def\uquark    {{\ensuremath{\Pu}}\xspace}
\def\uquarkbar {{\ensuremath{\overline \uquark}}\xspace}
\def\uubar     {{\ensuremath{\uquark\uquarkbar}}\xspace}
\def\dquark    {{\ensuremath{\Pd}}\xspace}
\def\dquarkbar {{\ensuremath{\overline \dquark}}\xspace}
\def\ddbar     {{\ensuremath{\dquark\dquarkbar}}\xspace}
\def\squark    {{\ensuremath{\Ps}}\xspace}
\def\squarkbar {{\ensuremath{\overline \squark}}\xspace}
\def\ssbar     {{\ensuremath{\squark\squarkbar}}\xspace}
\def\cquark    {{\ensuremath{\Pc}}\xspace}
\def\cquarkbar {{\ensuremath{\overline \cquark}}\xspace}
\def\ccbar     {{\ensuremath{\cquark\cquarkbar}}\xspace}
\def\bquark    {{\ensuremath{\Pb}}\xspace}

\def\tquark    {{\ensuremath{\Pt}}\xspace}

\def\pion   {{\ensuremath{\Ppi}}\xspace}
\def\piz    {{\ensuremath{\pion^0}}\xspace}
\def\pip    {{\ensuremath{\pion^+}}\xspace}
\def\pim    {{\ensuremath{\pion^-}}\xspace}
\def\pipm   {{\ensuremath{\pion^\pm}}\xspace}

\def\rhomeson {{\ensuremath{\Prho}}\xspace}
\def\rhoz     {{\ensuremath{\rhomeson^0}}\xspace}
\def\rhop     {{\ensuremath{\rhomeson^+}}\xspace}

\def\kaon    {{\ensuremath{\PK}}\xspace}

\def\KorKbar {\kern \thebaroffset\optbar{\kern -\thebaroffset \PK}{}\xspace}
\def\Kz      {{\ensuremath{\kaon^0}}\xspace}

\def\Kp      {{\ensuremath{\kaon^+}}\xspace}
\def\Km      {{\ensuremath{\kaon^-}}\xspace}
\def\Kpm     {{\ensuremath{\kaon^\pm}}\xspace}

\def\KS      {{\ensuremath{\kaon^0_{\mathrm{S}}}}\xspace}

\def\Dbar    {{\ensuremath{\offsetoverline{\PD}}}\xspace}
\def\D       {{\ensuremath{\PD}}\xspace}

\def\DorDbar {\kern \thebaroffset\optbar{\kern -\thebaroffset \PD}\xspace}
\def\Dz      {{\ensuremath{\D^0}}\xspace}
\def\Dzb     {{\ensuremath{\Dbar{}^0}}\xspace}
\def\Dp      {{\ensuremath{\D^+}}\xspace}
\def\Dm      {{\ensuremath{\D^-}}\xspace}

\def\DpDm    {\ensuremath{\Dp {\kern -0.16em \Dm}}\xspace}

\def\B       {{\ensuremath{\PB}}\xspace}
\def\Bbar    {{\ensuremath{\offsetoverline{\PB}}}\xspace}

\def\BorBbar {\kern \thebaroffset\optbar{\kern -\thebaroffset \PB}\xspace}
\def\Bz      {{\ensuremath{\B^0}}\xspace}
\def\Bzb     {{\ensuremath{\Bbar{}^0}}\xspace}
\def\Bd      {{\ensuremath{\B^0}}\xspace}

\def\BdorBdbar {\kern \thebaroffset\optbar{\kern -\thebaroffset \Bd}\xspace}
\def\Bu      {{\ensuremath{\B^+}}\xspace}
\def\Bub     {{\ensuremath{\B^-}}\xspace}
\def\Bp      {{\ensuremath{\Bu}}\xspace}
\def\Bm      {{\ensuremath{\Bub}}\xspace}
\def\Bpm     {{\ensuremath{\B^\pm}}\xspace}

\def\Bs      {{\ensuremath{\B^0_\squark}}\xspace}

\def\BsorBsbar {\kern \thebaroffset\optbar{\kern -\thebaroffset \Bs}\xspace}

\def\Bds     {{\ensuremath{\B_{(\squark)}^0}}\xspace}
\def\Bdsb    {{\ensuremath{\Bbar{}_{(\squark)}^0}}\xspace}

\def\jpsi     {{\ensuremath{{\PJ\mskip -3mu/\mskip -2mu\Ppsi}}}\xspace}

\def\Y#1S{\ensuremath{\PUpsilon{(#1S)}}\xspace}

\def\FourS {{\Y4S}}

\def\LorLbar     {\kern \thebaroffset\optbar{\kern -\thebaroffset \PLambda}\xspace}

\def\to                 {\ensuremath{\rightarrow}\xspace}

\def\CP                {{\ensuremath{C\!P}}\xspace}
\def\CPT               {{\ensuremath{C\!PT}}\xspace}

\def\Vud  {{\ensuremath{V_{\uquark\dquark}^{\phantom{\ast}}}}\xspace}
\def\Vcd  {{\ensuremath{V_{\cquark\dquark}^{\phantom{\ast}}}}\xspace}
\def\Vtd  {{\ensuremath{V_{\tquark\dquark}^{\phantom{\ast}}}}\xspace}

\def\Vubs  {{\ensuremath{V_{\uquark\bquark}^\ast}}\xspace}
\def\Vcbs  {{\ensuremath{V_{\cquark\bquark}^\ast}}\xspace}
\def\Vtbs  {{\ensuremath{V_{\tquark\bquark}^\ast}}\xspace}

\newcommand{\dmd}{{\ensuremath{\Delta m_{\dquark}}}\xspace}

\def\AT#1     {\ensuremath{A_{\mathrm{T}}^{#1}}\xspace}           

\def\C#1      {\ensuremath{\mathcal{C}_{#1}}\xspace}                       
\def\Cp#1     {\ensuremath{\mathcal{C}_{#1}^{'}}\xspace}                    
\def\Ceff#1   {\ensuremath{\mathcal{C}_{#1}^{\mathrm{(eff)}}}\xspace}        
\def\Cpeff#1  {\ensuremath{\mathcal{C}_{#1}^{'\mathrm{(eff)}}}\xspace}       
\def\Ope#1    {\ensuremath{\mathcal{O}_{#1}}\xspace}                       
\def\Opep#1   {\ensuremath{\mathcal{O}_{#1}^{'}}\xspace}                    

\newcommand{\aunit}[1]{\ensuremath{\text{\,#1}}}       

\newcommand{\tev}{\aunit{Te\kern -0.1em V}\xspace}
\newcommand{\gev}{\aunit{Ge\kern -0.1em V}\xspace}
\newcommand{\mev}{\aunit{Me\kern -0.1em V}\xspace}
\newcommand{\kev}{\aunit{ke\kern -0.1em V}\xspace}
\newcommand{\ev}{\aunit{e\kern -0.1em V}\xspace}
 
\newcommand{\mevc}{\ensuremath{\aunit{Me\kern -0.1em V\!/}c}\xspace}
\newcommand{\gevc}{\ensuremath{\aunit{Ge\kern -0.1em V\!/}c}\xspace}
\newcommand{\mevcc}{\ensuremath{\aunit{Me\kern -0.1em V\!/}c^2}\xspace}
\newcommand{\gevcc}{\ensuremath{\aunit{Ge\kern -0.1em V\!/}c^2}\xspace}
\newcommand{\gevgevcccc}{\ensuremath{\gev^2\!/c^4}\xspace} 

\def\fb   {\ensuremath{\aunit{fb}}\xspace}
\def\invfb   {\ensuremath{\fb^{-1}}\xspace}
\def\ab   {\ensuremath{\aunit{ab}}\xspace}
\def\invab   {\ensuremath{\ab^{-1}}\xspace}

\def\gsim{{~\raise.15em\hbox{$>$}\kern-.85em
          \lower.35em\hbox{$\sim$}~}\xspace}
\def\lsim{{~\raise.15em\hbox{$<$}\kern-.85em
          \lower.35em\hbox{$\sim$}~}\xspace}

\def\tell1  {TELL1\xspace}
\def\ukl1   {UKL1\xspace}

\begin{document}

\title{\CP violation in \B mesons}

\author{D.~Manuzzi\\
on behalf of the LHCb collaboration,\\with results from the Belle and Belle II collaborations.}
\affiliation{University of Bologna, via Irnerio 46, Bologna, 40126, BO, Italy}

\begin{abstract}
This proceeding reports a selection of recent experimental results concerning \CP violation in the sectors of \Bz and \Bpm mesons. 
They were published within the last two years by the \belle, \belletwo and \lhcb collaborations. The first set of measurements is related to the determination of the angles of the Unitarity Triangle. 
The second is connected to another crucial test of the SM: the isospin sum rule of $B\to K\pi$ decays. 
Thirdly, studies of \CP violation in the decays charged \B mesons to charmless three-body final states are presented.
Finally, prospects for the future experimental upgrades of this sector are summarised.
\end{abstract}

\maketitle

\thispagestyle{fancy}

\section{Introduction}
Several \CP-violating phenomena have been observed studying  \Bu and \Bd mesons.
Now, \CP violation is well established in the decays of both mesons~\cite{PDG2020}.
The \CP violation in the \Bz-\Bzb mixing has not been observed yet. 
However, the Standard Model (SM) predicts this effect to be suppressed below the current experimental limit~\cite{PDG2020,CPVmixingB}.
Instead,  the \CP violation in the interference between mixing and decay of neutral \B mesons is well measured~\cite{PDG2020}. 
These results strongly corroborate the SM and, in particular, the flavour-mixing mechanism of quarks encoded by the Cabibbo-Kobayashi-Maskawa matrix (CKM)~\cite{CKM}.
The investigation of this sector is a very active field. 
Indeed, the combination of multiple measurements provides crucial SM consistency tests, which are very sensitive to the eventual indirect effects of new physics (NP).
Some examples come from the relations called Unitarity Triangle (UT) and Isospin Sum Rule~\cite{PDG2020,Gronau:2005kz}.
Besides, the decays of \B mesons to more than two bodies have a rich \CP-violation phenomenology, whose connection with the underlying resonance contributions has to be scrutinised.
This proceeding reviews the most recent results on these topics. 
They have been obtained by the \belle, \belletwo, and \lhcb collaborations~\cite{Detectors}. 
Due to the large number of publications and the complexity of the analyses, only main aspects are highlighted here.
Finally, the programs for the future upgrades of  the \belletwo and \lhcb experiments are summarised, with a focus on the prospects for the measurements introduced above.

\section{Updates to the UT angles\label{sec:gamma}}
The SM assumes the unitarity of the CKM matrix. 
This condition implies various relations among its elements. 
One of them is $\Vud\Vubs + \Vcd\Vcbs + \Vtd\Vtbs = 0$.
It can be represented as a triangle in the complex plane, whose angles are~\footnote{Two notations are available in the literature. After defining them both, the one from the PDG~\cite{PDG2020}, namely $(\alpha,\beta,\gamma)$, is used.}
\begin{eqnarray*}
 \alpha 	&\equiv 	\phi_2	\equiv &	\arg\left[-(\Vtd\Vtbs)/(\Vud\Vubs)\right], \\ 
 \beta 		&\equiv 	\phi_1	\equiv &	\arg\left[-(\Vcd\Vcbs)/(\Vtd\Vtbs)\right], \\ 
 \gamma	&\equiv 	\phi_3	\equiv &	\arg\left[-(\Vud\Vubs)/(\Vcd\Vcbs)\right].
\end{eqnarray*}
Each of them can be determined either with direct measurements or indirectly through global fits of the CKM parameters,  assuming unitarity.
Any discrepancy between direct and indirect measurements, or between measurements exploiting different modes, would be a hint of NP.
The status of the art of these comparisons is reviewed by another talk at this conference~\cite{UTfit}.

Exploiting the interference between the Cabibbo-favoured $\bquark \to \cquark$ and the Cabibbo-suppressed $\bquark \to \uquark$ transitions, it is possible to measure $\gamma$ with processes dominated by tree-level Feynman diagrams.
This is usually done by studying $B^{\pm}\to (D\to f_D) h^\pm$ decays, where $h^\pm$ is a charged pion or kaon, and $f_D$ stands for a final state shared by the decays of both \Dz and \Dzb mesons.
In general, crucial parameters for the achievable precision are the ratio between the magnitudes of the amplitudes of the interfering modes ($r_\B$) and the difference between their strong phases ($\delta_\B$).
Depending on $f_D$, various methods are available in the literature. 
The GLW~\cite{GLW} method concerns the cases when $f_D$ is a \CP eigenstate ({\em e.g.} $f_D = \Kp\Km$).
The ADS~\cite{ADS} method exploits the interference between Cabibbo-favoured \Dz decays and doubly-Cabibbo-suppressed \Dzb decays ({\em e.g.} $f_D = \Kp\pim$).
Another option is using \Dz and \Dzb decays \CP self-conjugated multi-body final states ({\em  e.g.} $f_D = \KS h^+h^-$).
In this case,  the \D-decay phase space has to be considered.
A detailed description of the intermediate-resonance structure requires assumptions on the \D-decay amplitude model.
This implies potentially large and difficult-to-determine systematic uncertainties.
To overcome this issue a model-independent strategy, called BPGGSZ method~\cite{BPGGSZbinned}, was developed. 
It utilises \CP-asymmetry measurements in disjoint regions (bins) of the \D-decay phase space. 
Such measurements are then related to $\gamma$ using measurements of \D-decay parameters available in the literature~\cite{BESIII:DdecayParameters}.

The \belle and \belletwo collaborations have recently published a joint $\gamma$ determination following the BPGGSZ method.
The full decay chain is $\Bpm \to (D\to \KS h^+h^-) h^-$ where $h \in \lbrace \pi, K\rbrace$ and the integrated luminosity is $711\invfb$ from \belle plus $128\invfb$ from \belletwo.
The final result is~\cite{gammaBelleBelle2}:
\begin{eqnarray*}
\label{eq:res:gammaBelleBelle2}
\gamma &=& (78.4\pm 11.4\pm 0.5\pm1.0)^\circ,
\end{eqnarray*}
where the uncertainties are statistical, systematic and due to external inputs, respectively.
This is the most precise determination of $\gamma$ at the \B-factories. 
The improvement compared to the previous \belle results is equivalent to doubling the statistics.
This was possible thanks to the inclusion of the $D\to \KS\Kp\Km$ decay mode and improvements in the \KS selection and background suppression. 
A further decisive advantage is also taken from the inputs updated by \besiii~\cite{BESIII:DdecayParameters}.

\begin{figure}[b]
\centering
\includegraphics[width=0.35\textwidth]{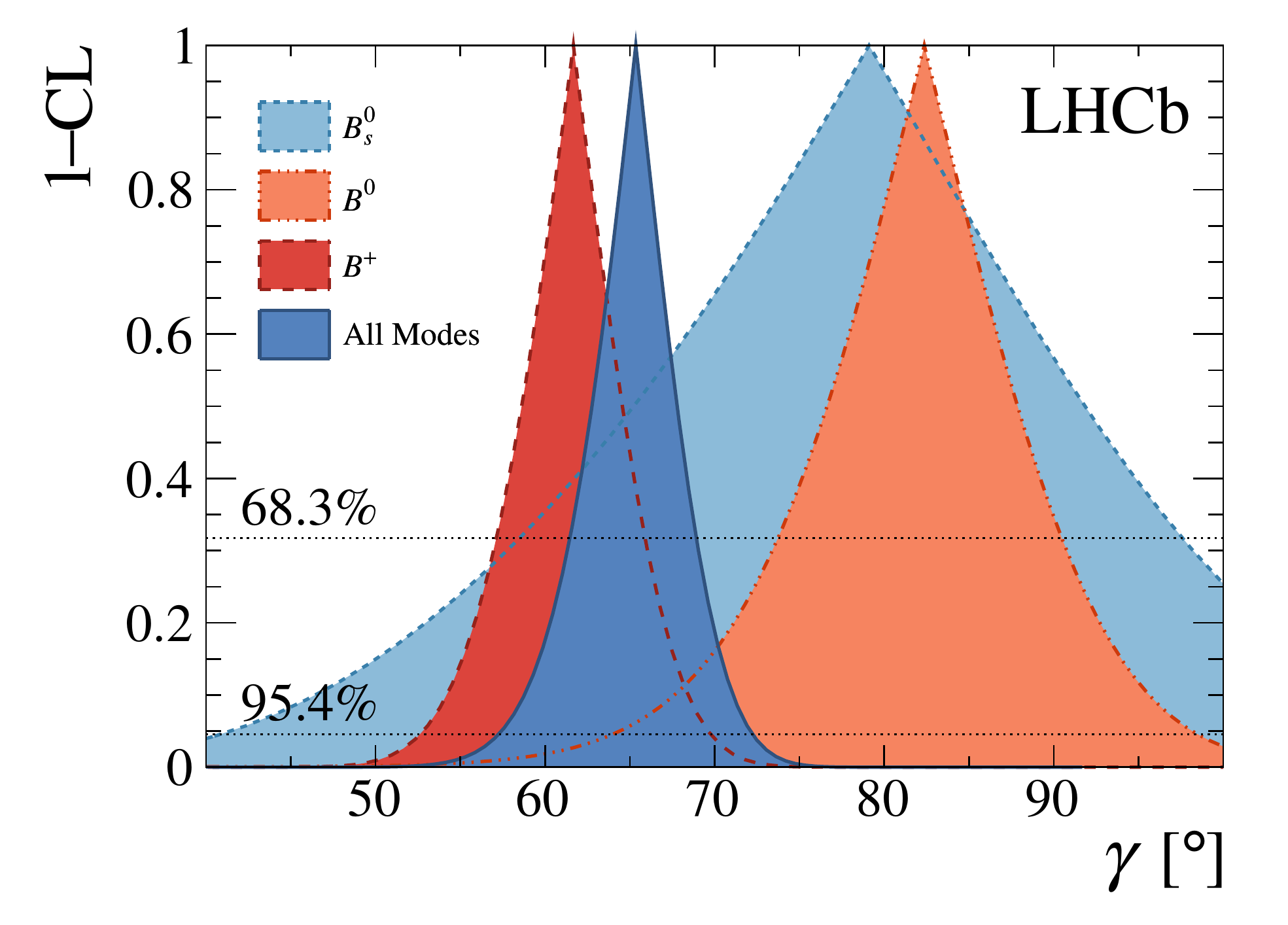}
\caption{One dimensional 1-CL profiles for the \lhcb $\gamma$ combinations using inputs from \Bs (light blue), \Bd (orange), \Bu mesons (red) and all species together (dark blue)~\cite{gammaLHCbComb}.}
\label{fig:gammaLHCbComb}
\end{figure}

In 2021 the \lhcb collaboration published the combination of all its direct $\gamma$ measurements.
It includes all the procedures cited above, but also methods based on the measurements of time-dependent decay rates to exploit the interference due to the mixing of \Bds and \Bdsb mesons.
The data were collected during the Run1 and Run2 of the \lhc, with a total integrated luminosity of 9\invfb. 
The combination follows a frequentist approach~\cite{gammaLHCbCombMethod}, with the novel inclusion of inputs from the \textit{charm} sector.
The final results is~\cite{gammaLHCbComb}:
\begin{eqnarray*}
\label{eq:res:gammaLHCbComb}
\gamma &=& (65.4^{+3.8}_{-4.2})^{\circ}.
\end{eqnarray*}
This is the most precise determination of $\gamma$ by a single experiment. 
It is statistically dominated and agrees with the indirect determinations from global CKM fits.
In Fig.~\ref{fig:gammaLHCbComb} the contributions to the \lhcb $\gamma$ combination coming from the \Bd, \Bu, and \Bs species are compared. 
A $2\sigma$ tension between the results from \Bd and \Bu sectors is observed.
The new combination method permitted the simultaneous determination the \Dz mixing parameters.
In particular, the precision on $y_D \equiv \Delta\Gamma/\Gamma$ was improved by a factor of two with respect to the previous world average.

Despite the release of this combination, the analysis of \lhcb data is still ongoing.
A more recent $\gamma$ measurement, concerns $\B^\pm \to (D\to h^+h^{\prime-}\piz)h^{\prime\prime\pm}$ decays, with $h, h^\prime, h^{\prime\prime} \in \lbrace \pi, K\rbrace$~\cite{gammaLHCbB2Dh}.
The case with $h = h^\prime$ are treated as ``quasi-GLW'', while the channels with $h\neq h^\prime$ are ``quasi-ADS''. 
In the first case, the $\D \to \pip\pim\piz$ and $\D\to \Kp\Km\piz$ final states are admixtures of \CP-even and \CP-odd eigenstates.
In the second case, the interference effects that are sensitive to $\gamma$ vary over the phase space of the \D decay due to the strong decays of intermediate resonances. 
To deal with that, dilution factors determined by measurements of \cleo-\cquark and \besiii~\cite{gammaLHCbB2DhInputs,BESIII:2021eud,Malde:2015mha} 
are exploited. 
This permits $\gamma$ to be calculated, combining eleven \CP observables described in Ref.~\cite{gammaLHCbB2Dh}. 
They are all measured with world best precision by this analysis.
The final results is~\cite{gammaLHCbB2Dh}
\begin{equation*}
\label{eq:res:gammaLHCbB2Dh}
\gamma = (56^{+24}_{-19})^\circ.
\end{equation*} 
It shall be used in future combinations to constrain the angle $\gamma$.
The total reported uncertainty is largely dominated by the statistical uncertainty.
In addition, the suppressed  $\Bp \to (D\to \pip\Km\piz) \Kp$  decay is observed for the first time, with a significance of almost $8\,\sigma$.

\begin{figure}
\begin{center}
\includegraphics[width=0.4\textwidth]{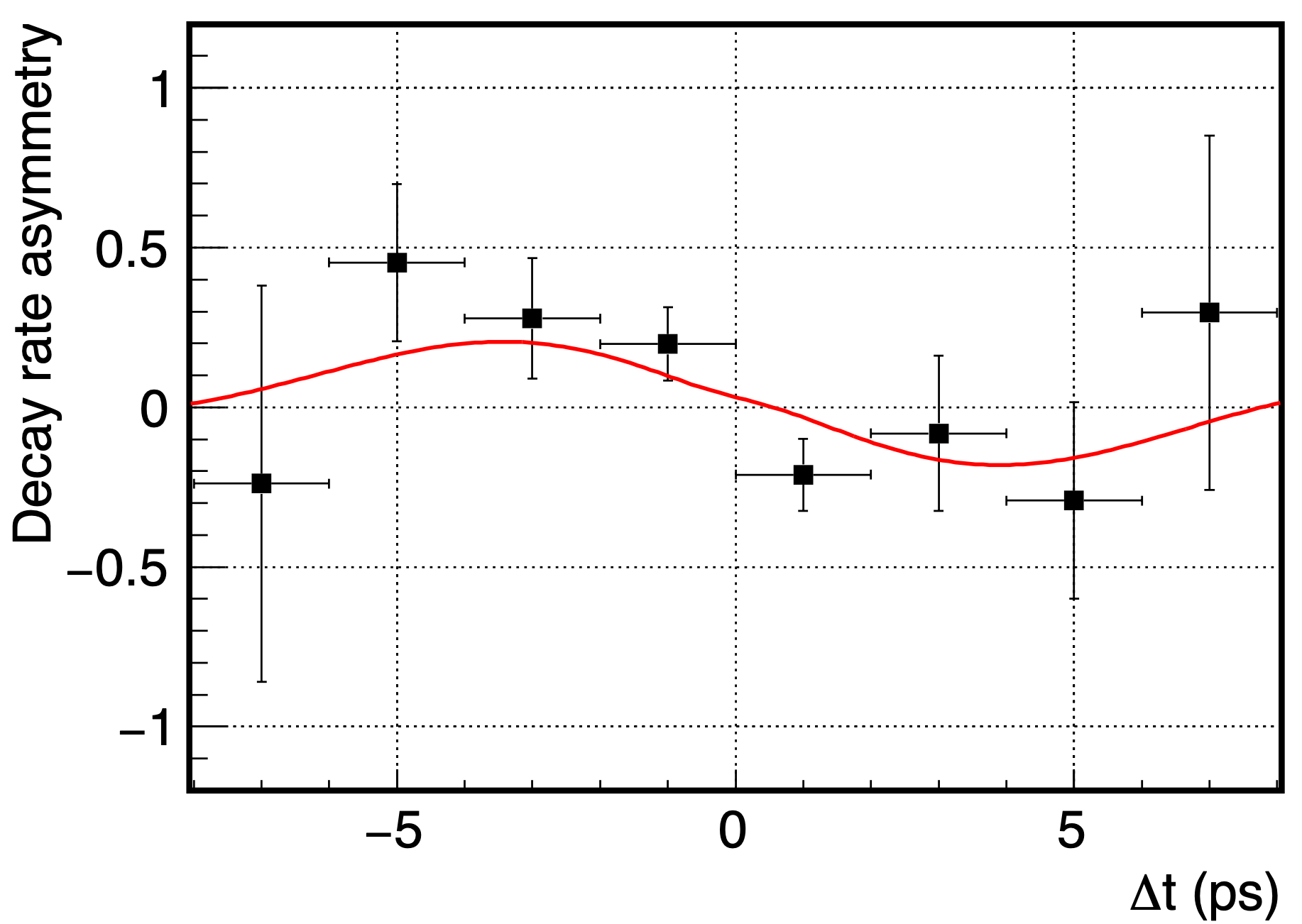}
\caption{Time dependent \CP asymmetry distribution obtained from $B^0\to \KS\KS\KS$ candidates (black dots) and curve coming from the result of the best fit (red line)~\cite{Belle:CPVB2KSKSKS}.}
\label{fig:B2KSKSKS}
\end{center}
\end{figure}

The UT angle $\beta$ can be measured using decays of neutral \B mesons to \CP eigenstates, $f_\CP$, in common between \Bz and \Bzb.
In these cases, assuming \CPT invariance and negligible \CP violation in the mixing, the time-dependent \CP symmetry can be written as\footnote{The notation for the time-dependent \CP asymmetries defined by the PDG is used here. Compared to the one used in Ref.~\cite{Belle:CPVB2KSKSKS}, it implies $C_f \!=\!-\mathcal{A}$.}
\begin{eqnarray*}
\tfrac{\Gamma_{\Bzb \to f_\CP}(t) - \Gamma_{\Bz \to f_\CP}(t)}{\Gamma_{\Bzb \to f_\CP}(t) + \Gamma_{\Bz \to f_\CP}(t)}
 &=&+S_{f_\CP} \sin (\dmd t)\\
 & &  -C_{f_\CP} \cos (\dmd t),
\end{eqnarray*}
where \dmd is the \Bz-\Bzb oscillation frequency, and the \CP violation in the decay and in the interference between mixing and decay are encoded by the parameters $C_{f_\CP}$ and $S_{f_\CP}$, respectively.
When only one CKM phase is present in the amplitudes which dominates the transition, the SM predicts $C_{f_\CP} = 0$ and $S_{f_\CP} = -\sin(2\beta)$.
The theoretically cleanest process is $\Bz\to\jpsi \KS$, which proceeds at tree level.
However, also channels dominated by a loop-level topology are possible.
They are particularly interesting because NP in the loops may have implications on the values of $C_{f_\CP}$ and  $S_{f_\CP}$.
The \belle collaboration recently studied with its full dataset (711 \invfb) one of these cases: the $B^0 \to \KS\KS\KS$ decay~\cite{belle:CPVB2KSKSKS:penguin}.
The analysis utilises a three-dimensional fit to extract signal and background yields.
The latter is mainly due to continuum background, namely  $\ep\en \to \uubar, \ddbar, \ssbar, \ccbar$ processes.
It is suppressed with a Neural Network classifier. 
The classifier output is also an observable of the fit.
The other ones are the beam-energy-constrained mass $M_{bc} = \sqrt{E^2_{\rm beam}- |\vec{p}_\B|^2}$ and the energy difference $\Delta E = E_\B - E_{\rm beam}$. 
The time-dependent \CP asymmetry is determined with a second fit.
Besides the signal-to-background fraction obtained in the first step, it use information about the flavour of the \B meson and its decay time.
Since neutral mesons produced by the process $\FourS\to \Bz\Bzb$ are selected,  the quark content of the signal \Bz (\Bzb)  can be tagged when a flavour-specific decay of the companion \Bzb (\Bz) is reconstructed.
The time information derives from the difference, $\Delta t$, between the decay times of the signal and tagging \B, measured in their center-of-mass frame.
Fig.~\ref{fig:B2KSKSKS} displays the time-dependent \CP asymmetry observed in data with the results of the best fit superimposed.
The determinations of this analysis are~\cite{Belle:CPVB2KSKSKS}
\begin{eqnarray*}
\label{eq:res:CPVB2KSKSKS}
S_{3\KS} &=& -0.71 \pm 0.23{\rm (stat.)} \pm 0.05{\rm (syst.)}, \\
C_{3\KS} &=&-\,0.12\pm 0.16{\rm (stat.)} \pm 0.05{\rm (syst.)}.
\end{eqnarray*}
They supersede the previous \belle results thanks to the improvements in the \KS selection, background suppression, vertex reconstruction, and increased statistics.
These values agree with the world average for $-\sin(2\beta)$~\cite{HFLAV2019} and the null SM prediction for $C_{3\KS}$. 
The significance of \CP violation in this mode is determined to be $2.5\,\sigma$ away from $(0, 0)$.

The measurement of the UT angle $\alpha$ relies on $\bquark\to \uquark\uquarkbar\dquark$ transitions. 
Due to the Cabibbo suppression of the tree-level topology, loop-level amplitudes may be sizeable.
Therefore, hadronic contributions, which are difficult to be precisely measured, complicate the determination of $\alpha$ from \B decays. 
A possible approach consists of combining measurements of decays related to isospin symmetries. 
The complete set of $\B \to \rho\rho$ isospin partners is valuable for this purpose~\cite{BelleII:B2rhorho:alphaMotivations}.
\belletwo is expected to be able to study them jointly and within the same experimental environment.
As a first step, it recently exploited the 190 \invfb  data collected so far to analyse the $\Bp \to \rhop\rhoz$ decay. 
Relevant inputs for the $\alpha$ determination are its branching ratio, \CP asymmetry, and longitudinal polarization fraction, $f_L$.
The whole reconstructed decay chain includes $\rhop \to \pip(\piz\to\gamma\gamma)$ and $\rhoz\to\pip\pim$ decays.
To measure $f_L$, the helicity angles between the momentum of the positive-charged pions and the direction opposite to the \Bp momentum are considered. 
The large continuum background is reduced with multivariate methods.
The final fit for the parameter extraction included 6D templates obtained from simulation and corrected with calibration samples, as described in Ref.~\cite{Belle-II:B2rhorho}.
The raw charge asymmetry determined by the fit is then corrected for the detection asymmetry due to instrumental effects. 
It is measured with a $\Dp\to \KS\pip$ control channel.
The final results are~\cite{Belle-II:B2rhorho}
\begin{eqnarray*}\label{eq:res:CPVB2rhoprho0}
A_\CP (\Bp \to  \rhop\rhoz) &=& -0.069 \pm 0.068  \pm 0.060, \\
\mathcal{B} (\Bp \to  \rhop\rhoz) &=& (23.2^{+2.2}_{-2.1} \pm 2.7)  \times 10^{-6},\\
f_L &=& 0.943^{+0.035}_{-0.033} \pm 0.027,
\end{eqnarray*}
where the first uncertainty is statistical and the second is systematic. 
The determinations are consistent with the measurements available in the literature~\cite{PDG2020}.
This is the first measurement of $A_\CP (\Bp \to  \rhop\rhoz)$ by \belletwo.
The systematic uncertainties are mainly data-driven, hence they are expected to decrease by exploiting the future datasets.

\section{Updates to the $\bm{K\pi}$ puzzle}
Isospin relations would suggest the same value for the \CP asymmetries of $\Bp\to \Kp\piz$ and $\Bz\to\Kp\pim$ decays. However, their world averages, recently updated by \lhcb~\cite{LHCb:CPV:B2Kpi}, show a $8\,\sigma$ discrepancy. 
This peculiarity constitutes the long-standing ``$K\pi$ puzzle''. 
As explained in Ref.~\cite{Gronau:2005kz}, a more accurate examination of this anomaly leads to the following sum rule, which also concerns the \CP asymmetries in the $\Bz\to\Kz\piz$ and $\Bp \to \Kz\pip$:
\begin{eqnarray*}
   		&A_{\CP}(B^0 \to K^+ \pi^-) 	&\mathcal{B}(B^0 \to K^+ \pi^-)		\tau_\Bp	  \\ 
+		&A_{\CP}(B^+ \to K^0 \pi^+)	&\mathcal{B}(B^+ \to K^0 \pi^+)	\tau_\Bz		\\ 
-	&2A_{\CP}(B^+ \to K^+ \pi^0) 	&\mathcal{B}(B^+ \to K^+ \pi^0)	\tau_\Bz		\\
-	&2	A_{\CP}(B^0 \to K^0 \pi^0) 	&\mathcal{B}(B^0 \to K^0 \pi^0)		\tau_\Bp	=0,
\end{eqnarray*}
where $\tau_\Bp$ ($\tau_\Bz$) is the lifetime of the \Bp (\Bz) meson.
Deviations above the $1\%$ level from this crucial condition are not allowed within the SM.
The observable with higher uncertainty is the \CP asymmetry of the $\Bz \to \Kz \piz$ decay.
\begin{figure}[b!]
\centering
\includegraphics[width=0.35\textwidth]{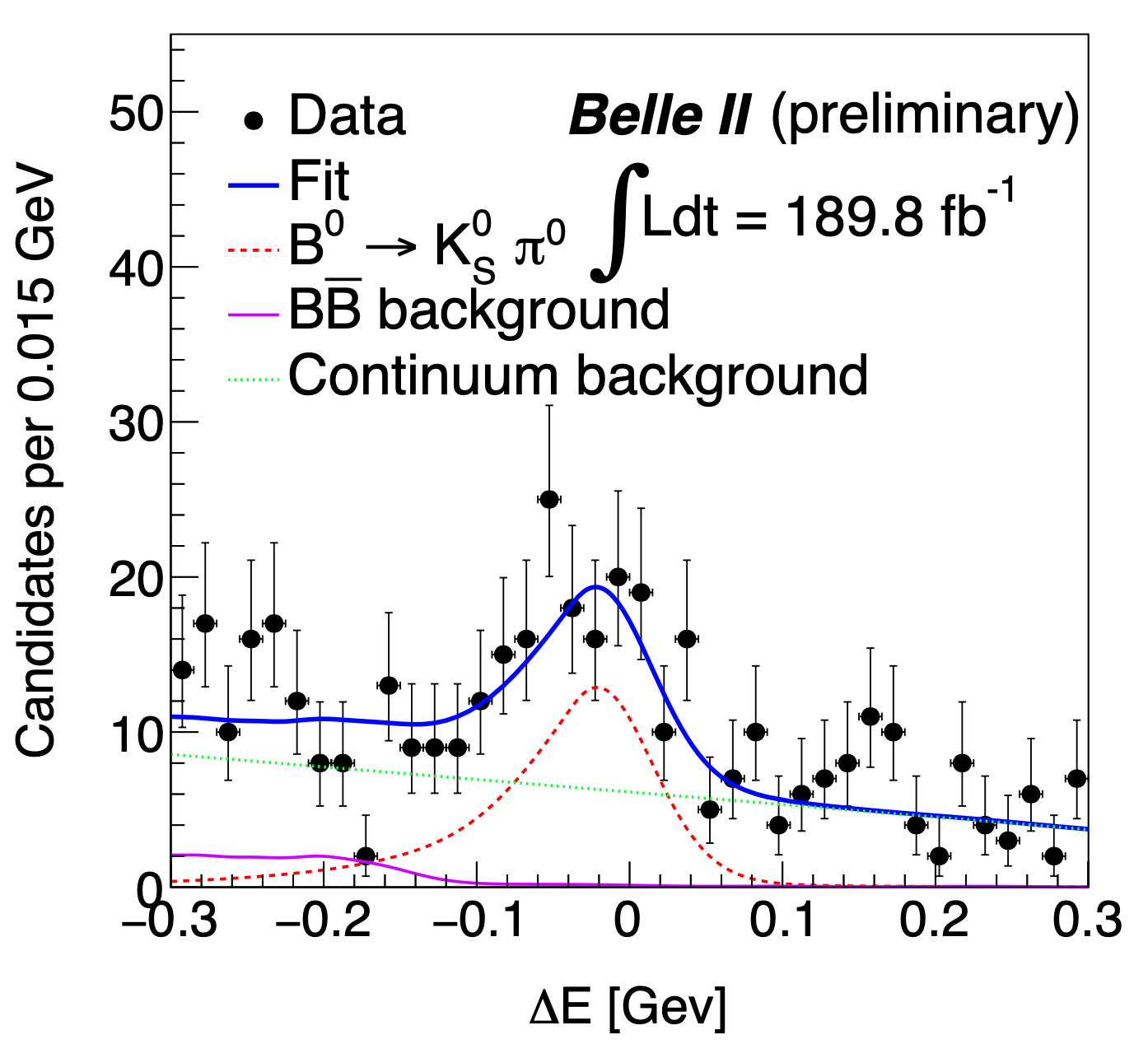}\\
\includegraphics[width=0.35\textwidth]{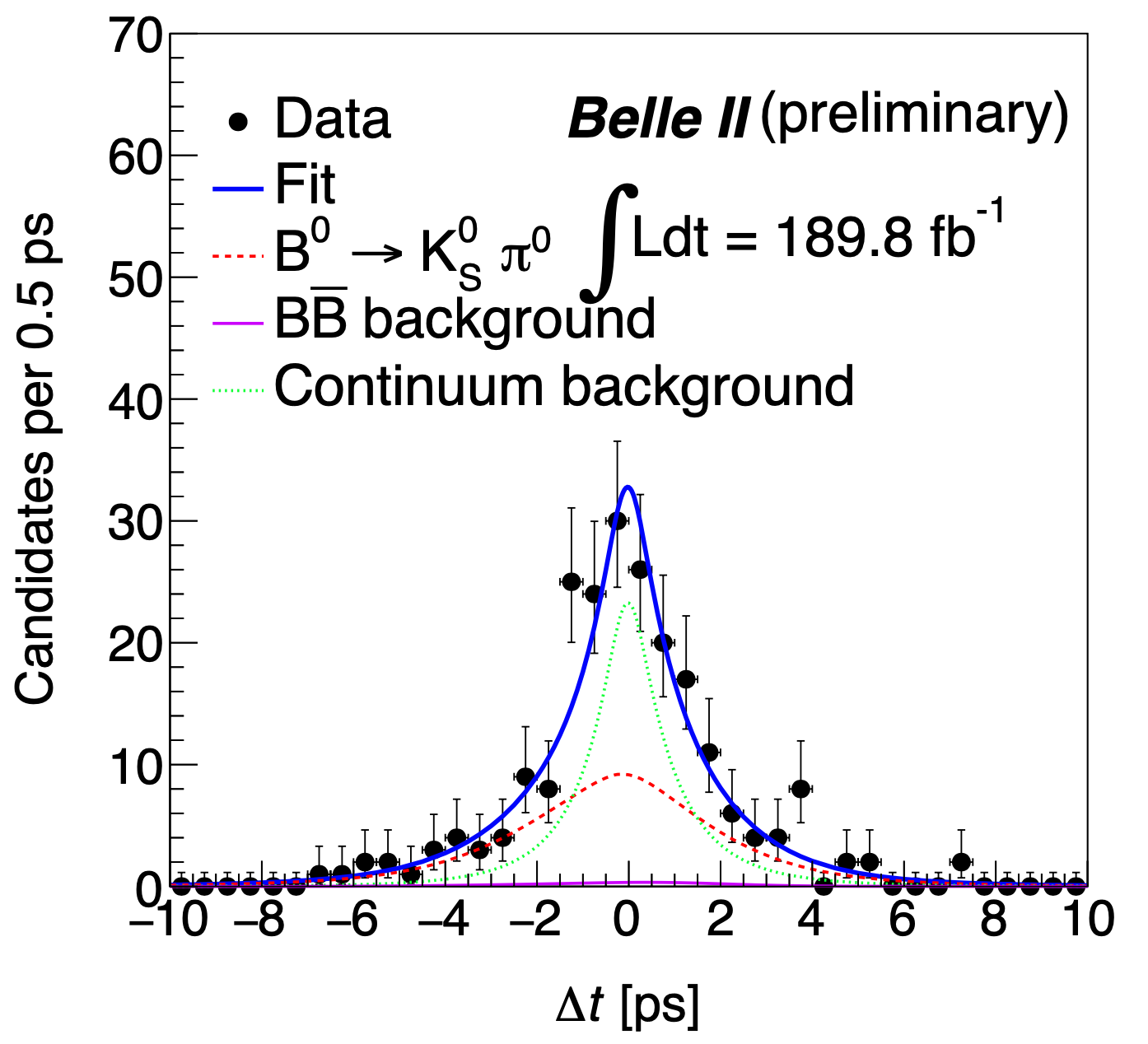}\\
\caption{Projections on the $\Delta E$ and $\Delta t$ axes of the fit used to determine $A_\CP(\Bz\to \Kz\piz)$.
More details in Ref.~\cite{Belle-II:B2KSpi0}.}
\label{fig:B2KSpi0}
\end{figure}
Currently, the sum-rule predicts  $A_{\CP}(\Bz \to \Kz \piz) = -0.138 \pm 0.025$, using the world averages of the other quantities~\cite{HFLAV2019}. 
Instead, the combination of the direct measurements by \belle and 	\babar returns  $A_{\CP}(\Bz \to \Kz \piz) = 0.01 \pm 0.10$, strongly requiring further inputs~\cite{PDG2020}.
The \belletwo collaboration recently updated this value using the 190 \invfb data collected so far.
The measurement analyses the decays of \B mesons to the \KS\piz final states.
The signal \B vertex is obtained by projecting back to the interaction region the flight direction of the \KS candidate.  This provides a good approximation of the signal \B decay vertex since both the transverse flight length of the \Bz meson and the transverse size of the interaction region are small compared to the \Bz flight length along the boost direction.
Another challenge is the suppression of the continuum background. 
A Boost Decision Tree algorithm is trained for this purpose.
The $\KS\piz$ final state cannot  be used to tag the \B flavour at the decay.
Therefore, the time-dependent \CP asymmetry is considered. 
The tagging information is obtained from the companion \B meson produced by the primary \FourS~resonance. 
The values of the $\tau_\Bz$, \dmd, and $S_{\KS\piz}$ are fixed to their world average to maximise the sensitivity on $A_{\CP}(\Bz \to \Kz \piz)$.
The $\Delta t$ resolution and bias are calibrated with $\Bz \to \jpsi \KS$ decays.
Fig.~\ref{fig:B2KSpi0} illustrates the projections of the fit on the $\Delta E$ and $\Delta t$ axes.
The final results are~\cite{Belle-II:B2KSpi0}
\begin{eqnarray*}
\label{eq:res:CPVB2KSpi0}
A_{\CP}(B^0\to K^0\pi^0) &=& -0.41^{+0.30}_{-0.32}\pm 0.09,\\
\mathcal{B}(B^0\to \Kz\pi^0) &=& (11.0\pm 1.2\pm 1.0)\times 10^{-6}
\end{eqnarray*}
This is the first determination of $A_{\CP}(B^0\to K_S^0\pi^0)$  using a time-dependent analysis at Belle II. 
The results agree with the previous measurements~\cite{HFLAV2019, Belle-II:B2KSpi0:timeIntegrated}.

\section{$\bm B$ decays to charmless 3-body final states}
\begin{figure}[]
\centering 
\includegraphics[width=0.37\textwidth]{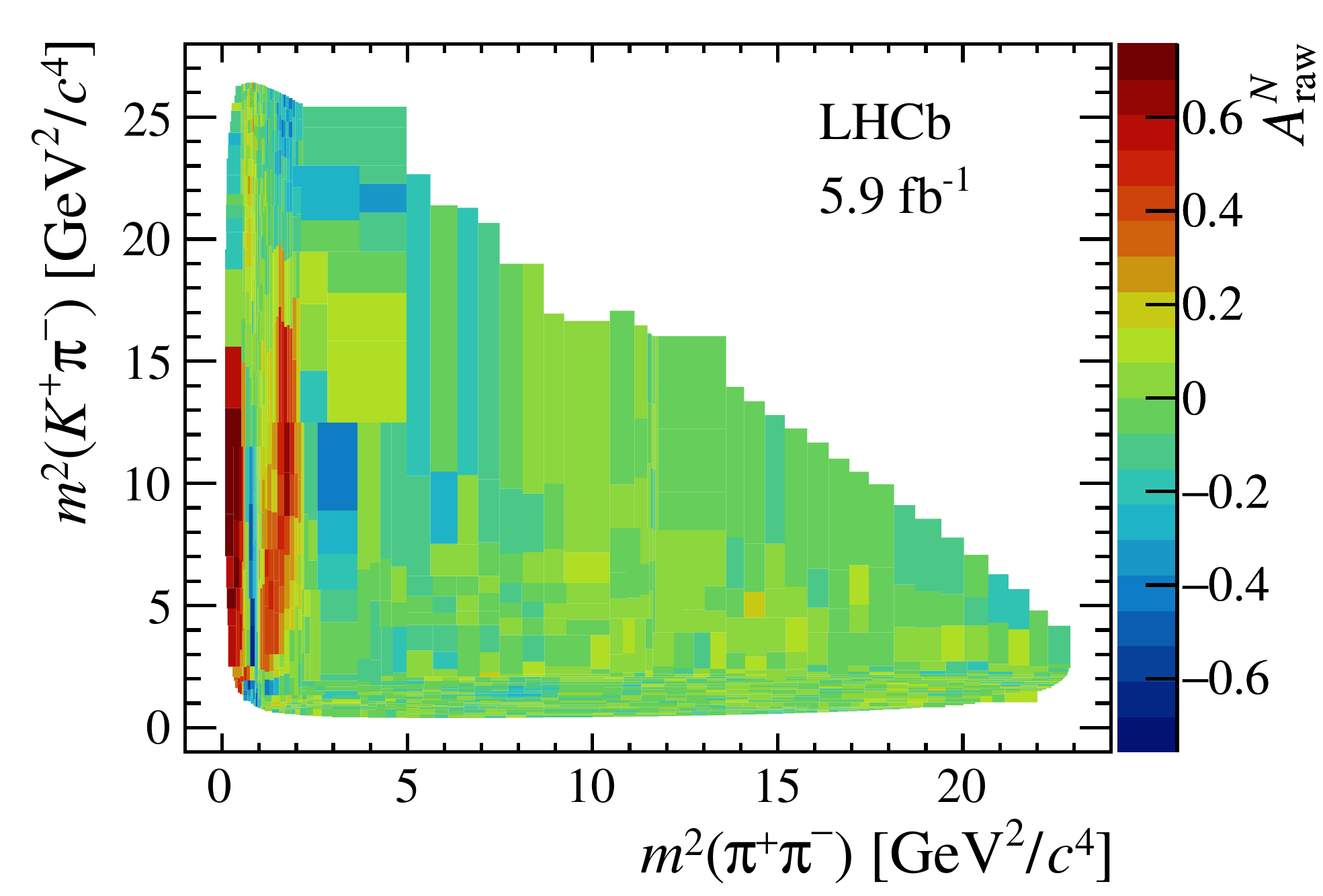}
\includegraphics[width=0.37\textwidth]{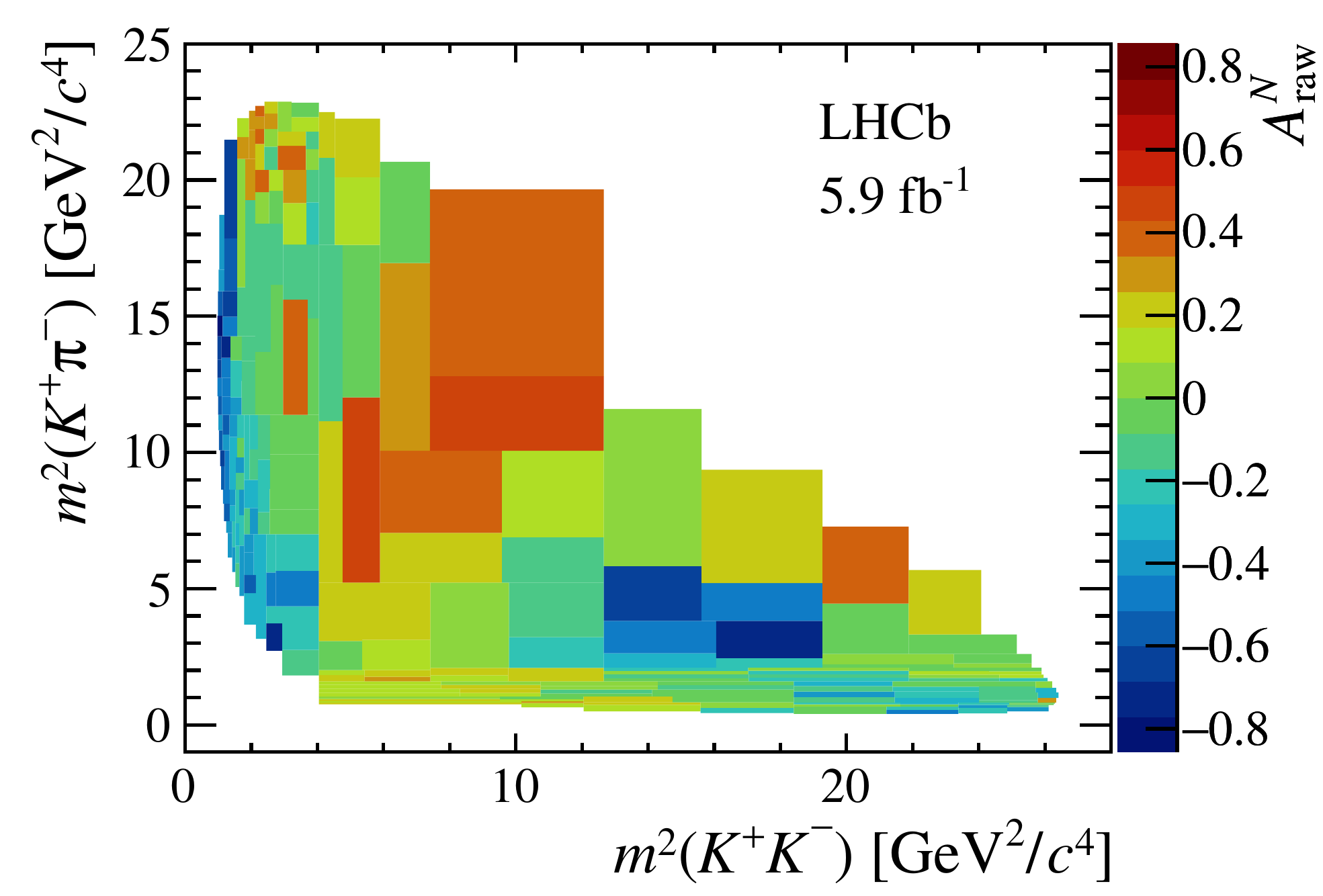}
\includegraphics[width=0.37\textwidth]{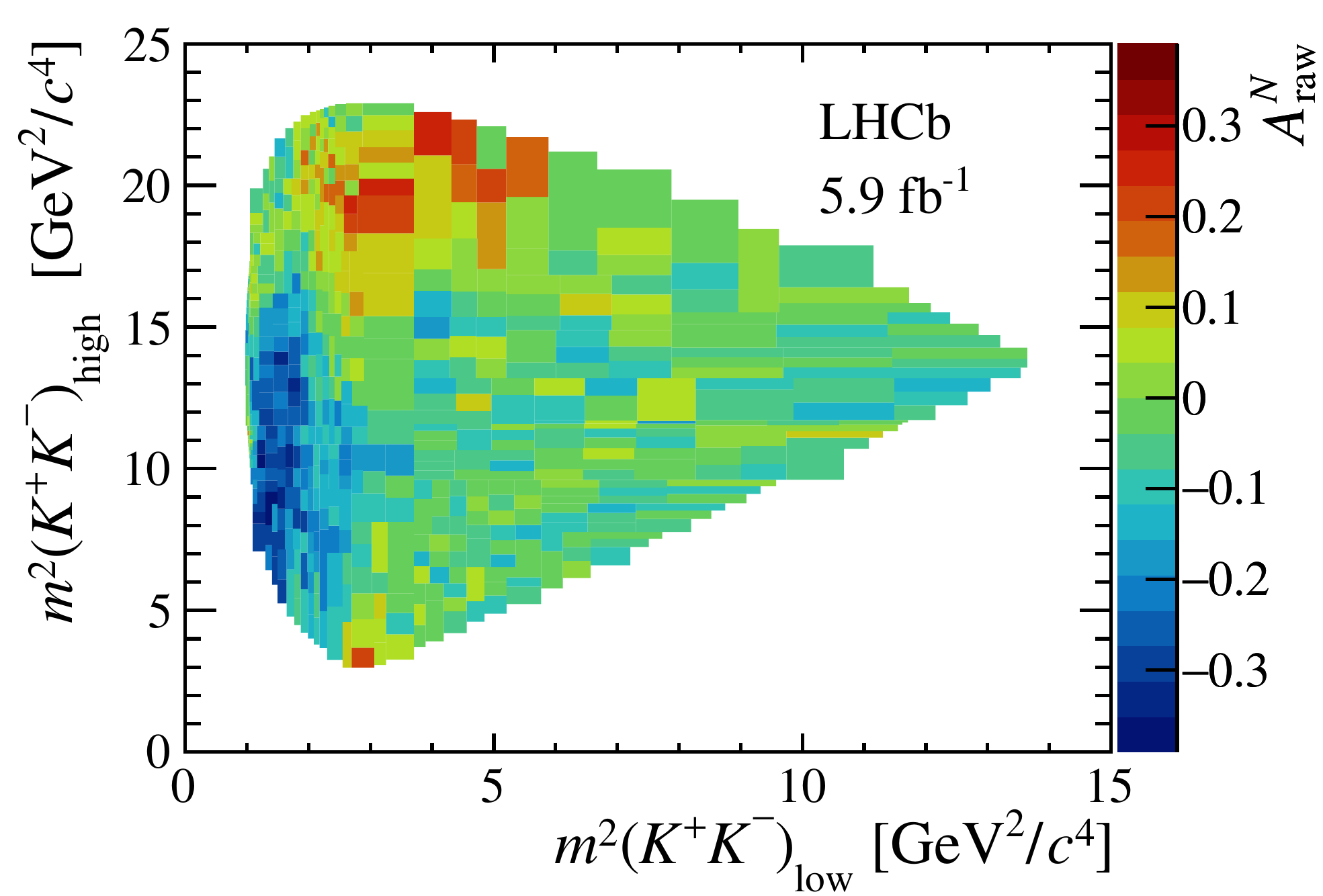}
\includegraphics[width=0.37\textwidth]{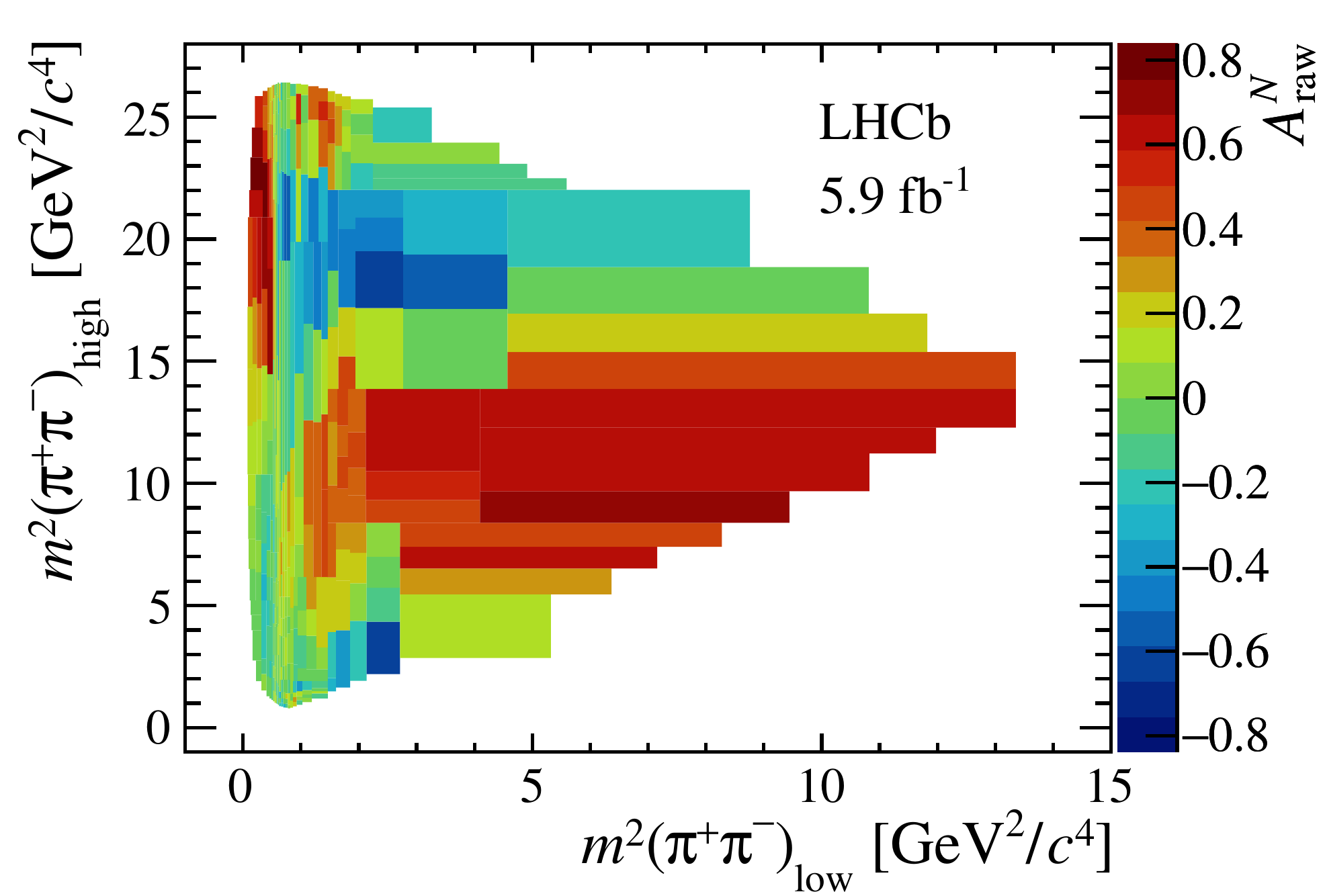}
\caption{\CP asymmetry distribution in bins of the Dalitz plot for $B^\pm\to K^\pm\pi^+\pi^-$ (top), $B^\pm\to \pi^\pm K^+K^-$ (middle-top),  $B^\pm\to K^\pm K^+K^-$ (middle-bottom), and  $B^\pm\to \pi^\pm\pi^+\pi^-$ (bottom) decays~\cite{LHCb:B2hhh}.}
\label{fig:B2hhh:CPVdalitz}
\end{figure}

A long-standing debate about the role of short- and long-distance contributions to the generation of the strong-phase differences, needed for direct \CP violation to occur, is ongoing in literature.
The three-body decays of \B mesons offer a way to solve it~\cite{Bediaga:2020qxg}.
Using Run1 data \lhcb showed evidence of global direct \CP-violation and high localised \CP asymmetries across the Dalitz plot in charmless 3-body \B decays~\cite{LHCb:2014mir}.
Amplitude analysis of $\Bpm \to \pipm\pip\pim$ connected the large \CP violation to the interference between S- and P-waves.
A similar analysis related the  $\pip\pim \leftrightarrow\Kp\Km$ rescattering~\cite{LHCb:B2hhh:prequels} to the large \CP violation observed in $\Bpm \to \pipm\Kp\Km$ decays.
\lhcb has just updated these results with the data collected during the Run2 (6\invfb).
The $\Bpm\to \Kpm\pip\pim$ and $\Bpm\to \Kpm\Kp\Km$ are also included.
The raw asymmetries between the yields of the \Bp and \Bm mesons are determined with a 
simultaneous fit to invariant mass spectra of the eight final states.
This permits the cross-feed backgrounds due to misidentification of the final state mesons to be better handled. 
This background source is particularly relevant because it peaks close to the signal.
After that, corrections for the experimental effect are applied.
The asymmetry due to the different detection efficiency of the charge conjugated final states is estimated with simulated candidates, calibrated with data-driven techniques.
The proton-proton collisions of the \lhc cause a slight difference between the production rates of \Bp and \Bm mesons.
This production asymmetry is estimated as the difference between the raw asymmetry of the $\Bpm \to \jpsi\Kpm$ decay and the world average of its \CP asymmetry, which is treated as an external input.
The global \CP asymmetries of the signal decay modes are determined to be~\cite{LHCb:B2hhh}
\begin{eqnarray*}
\label{eq:res:CPVB2hhh}
A_{C\!P}(B^\pm\to K^\pm\pi^+\pi^-)	&=& +0.011	\pm 0.002	\pm 0.003	\pm 0.003\\
A_{C\!P}(B^\pm\to \pi^\pm K^+K^-)    	&=& -0.144		\pm 0.007	\pm 0.003	\pm 0.003\\
A_{C\!P}(B^\pm\to K^\pm K^+K^-)    	&=& -0.037		\pm 0.002	\pm 0.002	\pm 0.003\\
A_{C\!P}(B^\pm\to \pi^\pm\pi^+\pi^-)	&=& +0.080	\pm 0.004	\pm 0.003	\pm 0.003
\end{eqnarray*}
where the uncertainties are statistical, systematic, and due to external inputs, respectively.
Large global \CP violation is measured for the latter three decay modes.
In the last two cases, it is observed for the first time.
Besides, the analysis was replicated in different regions (bins) of the Dalitz plot,
as shown in Fig.~\ref{fig:B2hhh:CPVdalitz}.
The \CP asymmetries are not uniformly distributed in the phase space, with positive and negative $A_\CP$ appearing in the same charged B decay channel.
The results of the previous LHCb analysis are confirmed.
Significant \CP violation is present for all the analysed channels in the $\pip\pim \leftrightarrow \Kp\Km$ rescattering region, namely between 1 and 2.25 \gevgevcccc in both \pip\pim and \Kp\Km invariant masses.
The combination of the present $A_\CP$ results with the recent \lhcb relative branching fraction results for these decays shows good agreement with the U-spin symmetry proposed in Ref.~\cite{LHCb:B2hhh:Uspin}.

\section{Prospects}
\begin{figure*}
\centering 
\includegraphics[width=0.33\textwidth]{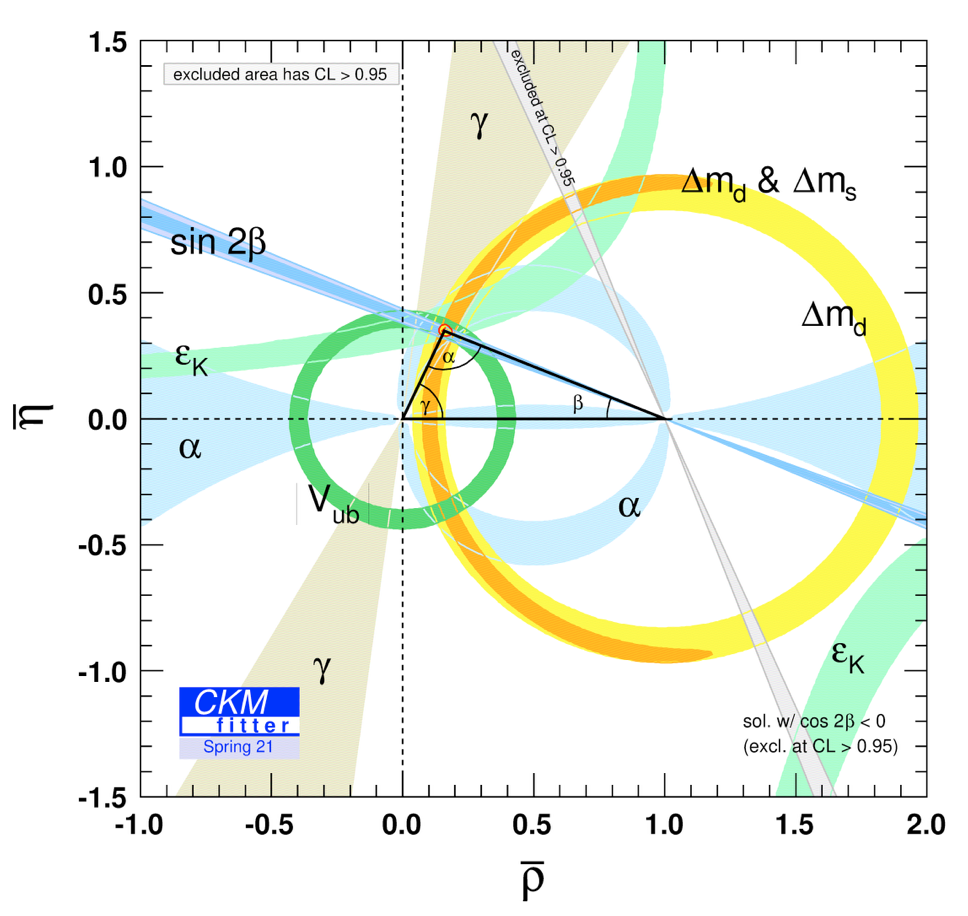}
\includegraphics[width=0.33\textwidth]{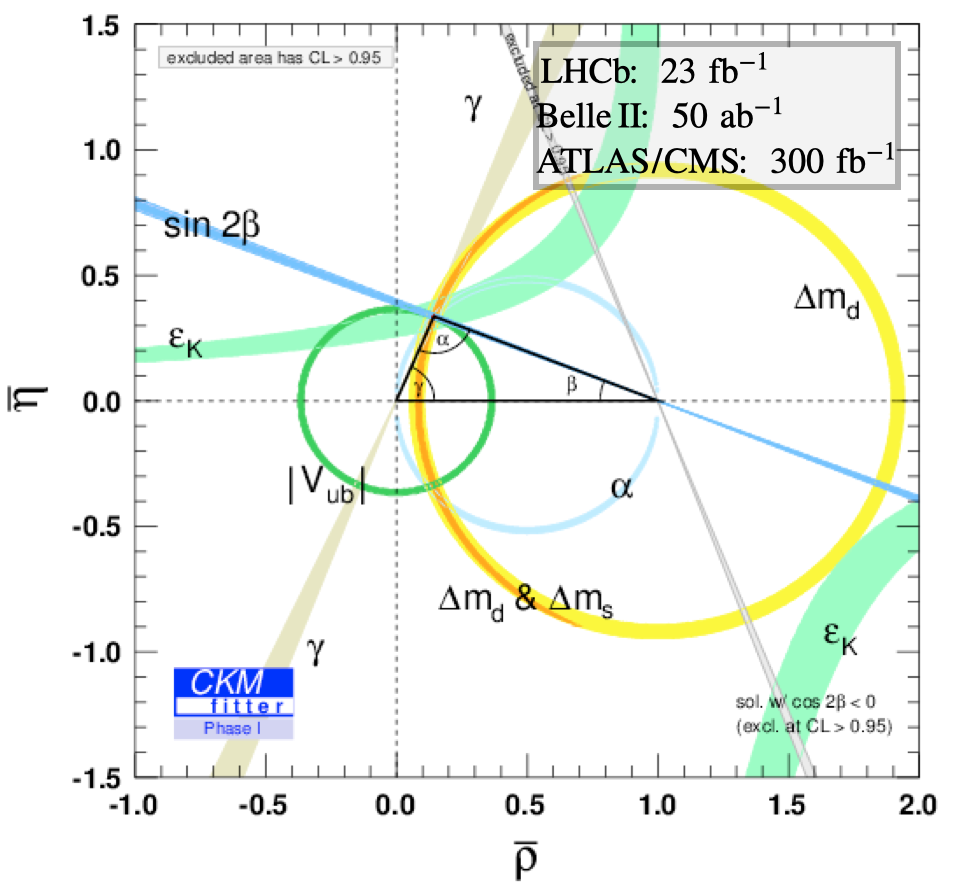}
\caption{Allowed parameter ranges for the UT parameters~\cite{CKMfitter}. 
The left plot is the current status of the art. The right plot assumes the current central values and includes the precision improvement foreseen for the future for both experiment and theory inputs.}
\label{fig:UTprospects}
\end{figure*}

Many of the \CP violation observables in the \B sector are dominated by statistical uncertainty. 
This is one of the motivations leading to the foreseen upgrades of the \lhcb and \belletwo detectors.
\lhcb is expected to increase its dataset to 23\invfb before the end of 2026, touching 50\invfb within 2032.
In the same period, the target of \belletwo is collecting a total of 50\invab.
Several studies have already started for the next phase, setting the \lhcb goal to 300\invfb and the \belletwo target to 250\invab.
Details about the crucial experimental challenges and the future physics programmes of both collaborations are documented in Ref.~\cite{Belle2:future,LHCb:future}.
Table~\ref{fig:UTprospects} summarises the expected uncertainties on the direct measurements of the UT angles.
\lhcb and \belletwo will provide comparable contributions to the determination of $\sin 2\beta$. 
\lhcb is expected to be more competitive for the $\gamma$ measurement, whereas \belletwo will achieve the most precise determinations of $\alpha$. 
The comparison between the current and future accuracy of the UT tests is illustrated in Fig.~\ref{fig:UTprospects} by the plots from the CKMfitter collaboration.

The further \CP violation features cited above will be also deeply investigated. 
\belletwo is expected to dominate the measurement of $A_{\CP}(\Bz \to \Kz \piz)$, testing the isospin sum rule at $3\%$ level with $50\invab$.
Instead, concerning \B decays to charmless three-body final states, \lhcb is expected to collect far larger yields with a much better signal-to-noise ratio.

In conclusion, this proceeding presented a selection of novel experimental results about the \CP violation in the \B sector.
They are just examples of the intense work ongoing in this field. 
It can provide powerful tests of the SM consistency.
They currently corroborate the SM, but a much higher level of precision is expected for the years to come.
This challenge will benefit from the competition and complementarity of the \lhcb and \belletwo physic programmes.

\begin{table}[b!]
\centering
\begin{tabular}{l|ccc}
\hline
\textbf{Dataset} 			& $\bm{\alpha}$ 		& $\bm{\sin 2\beta}$ 		& $\bm{\gamma}$ \\
\hline
Current world average & $4.8^\circ$	& $0.017$ 		& $3.5^\circ$\\
\hline
\lhcb, 23\invfb			&						&	$0.006$		&	$1.5^\circ$ \\
\belletwo, 50\invab		& 	$0.6^\circ$	&	$0.005$		&	$1.5^\circ$ \\
\hline
\lhcb, 300\invfb			&						&	$0.003$		&	$0.35^\circ$ \\
\belletwo, 250\invab	& 	$0.3^\circ$	&	$0.002$		&	$0.8^\circ$ \\

\hline
\end{tabular}
\caption{Current and expected uncertainty on the direct determinations of the UT angles~\cite{Belle2:future,LHCb:future,HFLAV2019}.}
\label{tab:UTprospects}
\end{table}

\end{document}